# Algorithm of quantum engineering of large-amplitude high-fidelity Schrödinger cat states in setup with $k$ beam splitters and with inefficient photon number resolving detection

**Mikhail S. Podoshvedov[1,2], Sergey A. Podoshvedov[1*] and Sergei P. Kulik[1,3]**

[1] *Laboratory of quantum engineering of light, South Ural State University (SUSU), Chelyabinsk, Russia*
[2] *Institute of Physics, Kazan Federal University (KFU), Kazan, Russia*
[3] *Quantum Technology Centre, M.V. Lomonosov Moscow State University, Moscow, Russia*
[*] sapodo68@gmail.com

**Abstract**
We present an algorithm of quantum engineering of large-amplitude≥ 5 high-fidelity≥ 0.99 even/odd Schrödinger cat states (SCSs) using a single mode squeezed vacuum (SMSV) state as resource. Set of $k$ beam splitters (BSs) with arbitrary transmittance and reflectance coefficients sequentially following each other acts as a hub that redirects a multiphoton state into the measuring modes simultaneously measured by photon number resolving (PNR) detectors. We show that the multiphoton state splitting guarantees significant increase of the success probability of the SCSs generator compared to its implementation in a single PNR detector version and imposes less requirements on ideal PNR detectors. We prove that the fidelity of the output SCSs and its success probability are in conflict with each other (which can be quantified) in a scheme with ineffective PNR detectors, especially when subtracting large (say, 100) number of photons, i.e., increasing the fidelity to perfect values leads to a sharp decrease in the success probability. In general, the strategy of subtracting up to 20 photons from initial SMSV in setup with two BSs is acceptable for achieving sufficiently high values of the fidelity and success probability at the output of the generator of the SCSs of amplitude≤ 3 with two inefficient PNR detectors.

**Introduction**
In quantum optics the notion of the superposition of classically distinguishable macroscopic states[1] finds its embodiment in a form of superposition of coherent states with amplitudes equal in magnitude but opposite in sign[2,3]. Macroscopic SCSs can be of great significance in the demonstration of the fundamental problems[4,5]. In addition, the superpositions are well adjusted for the implementation of the quantum protocols[6-13]. The interaction of two coherent states with amplitudes equal in magnitude on a balanced beam splitter (BBS) guarantees an increase in the modulus of one coherent state by $\sqrt{2}$ times, while, at the same time, leaving the other output mode in the vacuum state $(|\beta\rangle|\beta\rangle \rightarrow |\sqrt{2}\beta\rangle|0\rangle)$. Furthermore, mixing of the components of entangled coherent states (ECSs) on the BBS separates them on their parity assuming that the vacuum state is an even state[6,7,9]. This allows for nearly deterministic implementation of Bell state measurement (BSM) by means of two PNR detectors. To reduce the contribution of an event when both PNR detectors are silent which may occur, SCSs of amplitude $\beta \geq 2$ are required to provide the sufficiently high degree of orthogonality[9] of the coherent states $|\pm\beta\rangle$. As quantum protocols work on the principle of producing the required state with subsequent measurement to obtain information stored in the prepared state, it is not surprising that the need to put into practice high-amplitude SCSs as well as entangled both with each other and with the photonic states (hybrid states) provoked a fairly large drive[14-32]. ECSs can be realized by passing pure SCSs through the BBS. Development of advanced technologies of quantum engineering of nonclassical states can be used to implement more complex entangled states with more than two distributed coherent nodes.

Practical realization of the optical SCSs mainly relies on nondeterministic photon subtraction technique, being the key for quantum engineering of the nonclassical CV states. The first implementation of the technique is based on passage of a Gaussian quantum state through highly transmitting beam splitters[14] (HTBS). By detecting photons in the measurement channel, catlike state may be generated in the output channel. But the detection of photons diverted by such HTBS into the measuring mode becomes a fairly rare event even in the case of small values of the SCS amplitude $\beta \leq 2$. Undoubtedly, given the serious progress in PNR detection based on transition-edge sensor (TES) technology[33-36], further modification of the photon subtraction technique could go in the direction of increase of number of subtracted photons[17,18,20,21,32], as well as using BS with arbitrary parameters[29-32] to have a chance to enhance output state characteristics. So, TES detector in[36] has a near unity detection efficiency and feature resolution below 20 photons at 1535 $nm$ but can detect up to 100 photons with a few-photon uncertainty which makes it appropriate to take theoretical consideration of subtracting one hundred and even more photons from the initial Gaussian SMSV state. In general, despite the large number of proposals and demonstrations of the SCS prototypes[16,18,19,21,26-28], the problem of generating large-amplitude with $\beta \geq 2$ even/odd SCS states remains unresolved which hinders development of quantum protocols with coherent states of light. Here, we use a hub of $k$ sequentially arranged BSs with arbitrary transmission and reflection amplitudes in each measuring mode which PNR detector is set to generate measurement-induced even/odd SCSs. As a component from which the target state is generated, the SMSV is applied as a state that is routinely used in practice. Moreover, use of the SMSV as input to the hub is quite natural as the state has even parity as well as the even SCS and can approximate it with sufficiently high fidelity for small amplitudes $\beta \leq 1$[23]. First, we introduce a family of CV states of a certain parity realized by the hub by means of subtraction of arbitrary number of photons from the SMSV state. The parity of the number of photons extracted from the SMSV in an indistinguishable manner determines the parity of the generated states. We show the family of the CV states of certain parity depends on one parameter. Second, we numerically demonstrate a possibility of generating even/odd SCSs amplitudes $\beta \geq 5$ with a fidelity exceeding $> 0.99$ (so-called perfect values) at the exit from the hub in the case of subtraction of 90,91 photons from the initial SMSV by ideal PNR detectors. Third, we show advantage of using multiphoton state demultiplexing resulting in a significant gain in the success probability of generating the SCSs on compared to the case of redirecting the multiphoton state into one PNR detector. In addition, the use of several PNR detectors reduces the requirements on maximum number of measurable photons detected with single-photon resolution. Fourth, we show that the characteristics (fidelity and success probability) of the output SCSs compete with each other in the practical case of using imperfect PNR detectors, which to a large extent can prevent the creation of a larger amplitude SCSs generator when it is required to subtract large number (say, 100) of photons. To avoid the competition, it is required to extract a smaller (say, 20) number of photons to guarantee obtaining close-to-ideal values of the output parameters.

**Results**
**Perfect values of even/odd SCSs parameters at the hub exit.** Let us consider the passage of SMSV state through a system of $k > 0$ lossless beam splitters BSs ($BS_i$) with real transmittance $t_i > 0$ and reflectance $r_i > 0$ coefficients ($i = 1, ..., k$) satisfying the normalization condition $t_i^2 + r_i^2 = 1$ and arranged in a row one after another, as shown in figure 1. No other states (vacuum) are applied to the second input of each beam splitter. In figure 1, the SMSV state occupies zero mode while the auxiliary modes denoted by $i$ ($i = 1,2,...,k$) are terminated by $k$ PNR detectors. The original SMSV state is given by

$$|SMSV\rangle = \frac{1}{\sqrt{\cosh s}} \sum_{n=0}^{\infty} \frac{y_0^n}{\sqrt{(2n)!}} \frac{(2n)!}{n!} |2n\rangle, \qquad (1)$$

where the parameter $s > 0$ is the squeezing amplitude of the SMSV state defining amount $y_0 = \tanh s/2 \leq 0.5 \geq 0$. It follows from formula (1) the SMSV state is described by two parameters $s$ and $y_0$, respectively. Furthermore, the input state (1) can also be characterized by two more parameters, namely, the squeezing $S$ expressed in $dB$ as $S = -10 \log(exp(-2s))$ and the mean number of photons $\langle n \rangle_{SMSV} = \sinh^2 s$ in the state. The absence of input SMSV (vacuum) is determined by the values $s = 0$, $y_0 = 0$, while the value of $y_0 = 0.5$ corresponds to the non-physical case of an infinitely large squeezing amplitude $s \to \infty$ of the original SMSV. The optical scheme in Fig. 1 can be considered as a hub composing of $k$ elements. Each $i$-element of the hub consists of one BS and PNR detector capable of resolving the number of photons to within one redirected by the BS from the initial SMSV state.

Subtraction of total either even $N_k = n_1 + n_2 + \cdots + n_k = 2m_N$ or odd $N_k = n_1 + n_2 + \cdots + n_k = 2m_N + 1$ number of photons from original SMSV in an indistinguishable manner so all information about which Fock state of the original superposition the photons are subtracted is lost generates the whole family of heralded CV states of definite parity in Eqs. (23,24). The projection of the entangled state in Eq. (22) onto Fock states can be realized by the simultaneous registration of $n_1 ..., n_i, ... n_k$ photons in $k$ measurement modes by modern TES detectors resolution of which has been improved[32,33,35]. The measurement-induced CV states have a well-defined parity either even (the superpositions in Eq. (23) involve only even number states) or odd (the CV states in Eq. (24) consist exclusively of odd number states) in dependency on the parity of the measurement outcomes. Therefore, the family can be divided into two subfamilies by its parity, namely, even (Eq. (23)) and odd (Eq. (24)), respectively. As the state in Eq. (23) becomes SMSV state in Eq. (1) in the case of $m_N = 0$, the SMSV state belongs to an even subfamily of the family of the CV states of definite parity. The success probabilities to realize the $2m_N, 2m_N + 1$ CV states follow from definition of the amplitudes of the entangled state in Eq. (25): $P_{n_1,n_2,...,n_k}^{(0,0,...,0)} = \left|C_{n_1,n_2,...,n_k}^{(0,0,...,0)}\right|^2 / \cosh s$ with the normalization condition $\sum_{n_1,...,n_i,...,n_k=0}^{\infty} P_{n_1,n_2,...,n_k}^{(0,0,...,0)} = 1$.

The optical scheme in Fig. 1 can be used for quantum engineering of measurement-induced even/odd SCSs. Indeed, such a task makes sense since the heralded CV states of definite parity may have photon distributions similar to even/odd SCSs ones which are given by

$$|SCS_+\rangle = 2N_+(\beta) exp(-\beta^2/2) \sum_{n=0}^{\infty} \frac{\beta^{2n}}{\sqrt{(2n)!}} |2n\rangle, \qquad (2)$$

$$|SCS_-\rangle = 2N_-(\beta) exp(-\beta^2/2) \sum_{n=0}^{\infty} \frac{\beta^{2n+1}}{\sqrt{(2n+1)!}} |2n+1\rangle, \qquad (3)$$

where $N_\pm = \left(2(1 \pm exp(-2\beta^2))\right)^{-1/2}$ are the corresponding normalization factors and $\beta > 0$ is an amplitude of the SCSs. For example, the difference between the heralded and target states is in that the amplitude $\beta$ to the power of $2m$ ($\beta^{2m}$) in Eq. (2) unlike the parameter $y_k$ to the power of $m$ ($y_k^m$) in Eq. (23). Instead of the difference between $\beta^{2m}$ and $y_k^m$, the CV states in Eq. (23) have an additional factor either $(2(n+m_N))!/(n+m_N)!$ associated with the number of extracted photons which can compensate for the difference between $\beta^{2m}$ and $y_k^m$. To estimate how close the measurement induced $2m_N, 2m_N + 1$ CV states can be to the target even/odd SCSs, one uses parameter fidelity

$F_{2m_N} = \left|\left\langle SCS_+ \middle| \Psi_{2m_N}^{(0,0,\ldots,0)} \right\rangle\right|^2$ and $F_{2m_N+1} = \left|\left\langle SCS_- \middle| \Psi_{2m_N+1}^{(0,0,\ldots,0)} \right\rangle\right|^2$ for two pure states. Ideal fidelity $F_{2m_N,max} = F_{2m_N+1,max} = 1$ can indicate on identity of the two states $|\Psi_{2m_N}^{(0,0,\ldots,0)}\rangle = |SCS_+\rangle$ and $|\Psi_{2m_N+1}^{(0,0,\ldots,0)}\rangle = |SCS_-\rangle$. We are interested in finding such conditions that provide the highest possible value of the fidelity, which nevertheless is less than one. In what follows, we do not use the subscript $max$ for the highest possible fidelity.

The dependences of the maximum possible fidelities $F_{2m_N}, F_{2m_N+1}$ of the CV states in Eqs. (23,24) on the SCS amplitude $\beta$ are shown in Figure 2. The number of subtracted photons varies from 0 to 90 for even (Figs. 2(a,b)) and from 1 to 91 (Figs. 2(c,d)) for odd CV states. In general, the more photons are subtracted from the initial SMSV, the higher the fidelity of the generated even/odd SCSs with greater amplitude $\beta$. So, in the case of extracting more than 80 photons from original SMSV, even/odd SCSs of amplitude of more than 5 can be generated with fidelity exceeding 0.99. The maximum possible fidelity $F_{2m_N}, F_{2m_N+1}$ solely depends on one parameter either $y_{2m_N}(\beta)$ or $y_{2m_N+1}(\beta)$. The parameters $y_{2m_N}(\beta)$ and $y_{2m_N+1}(\beta)$ that provide the fidelity values in Figs. 2 are shown in Figure 3 in dependency on the SCS amplitude $\beta$. The obtained values of $y_{2m_N}(\beta)$ and $y_{2m_N+1}(\beta)$ can be used to select the BS transmittances $t_i$ and also the squeezing amplitude $s$ of the initial SMSV state. Note that the reverse rule is observed for values of $y_{2m_N}(\beta)$ and $y_{2m_N+1}(\beta)$. The more photons are extracted from SMSV state, the smaller the value of the parameter $y_{2m_N}(\beta)$ and $y_{2m_N+1}(\beta)$ is required to generate the target even/odd SCSs. If the value of the parameter $y_k$ becomes less of either $y_{2m_N}(\beta)$ $(y_k < y_{2m_N}(\beta))$ or $y_{2m_N+1}(\beta)$ $(y_k < y_{2m_N+1}(\beta))$, then the generation of the even/odd SCSs with the fidelity shown in Figure 2 is impossible for given value of $\beta$. This imposes certain restrictions on the squeezing amplitude $s$ of the original SMSV state and, as a consequence, on the value of the initial parameter $y_0$ taking into account the fact that the passage of the SMSV state through the BSs reduces the value of $y_0$. So, in the case of using a system of one BS and PNR detector, initial value $y_0^{(1)}$ (here the superscript indicates a hub with the corresponding number of elements) of SMSV must meet the condition $y_0^{(1)} \geq y_{2m_N,2m_N+1}(\beta)$ to provide required fidelity in Figure 2 for a preselected range of $\beta$. If a system with $k$ BSs and PNR detectors is used in Fig. 1, then the following stronger constraint on $y_0^{(k)}$, i.e. $y_0^{(k)} \geq y_{2m_N,2m_N+1}(\beta)$, must be imposed to guarantee the fidelity shown in Figure 2 for required values of $\beta$. The inequality can imply use of the SMSV state with larger squeezing amplitude $s$ to provide the more value of $y_0^{(k)}$ on compared with $y_0^{(1)}$, i.e. $y_0^{(k)} > y_0^{(1)}$. It is also interesting to note a rather strong density of curves $y_{2m_N}(\beta)$ and $y_{2m_N+1}(\beta)$ at certain values of the SCS amplitude $\beta$ in Figures 3, especially, in the range of $0.11 \geq y_{2m_N,2m_N+1}(\beta) > 0$. This may mean that at a certain value of either $y_k = y_{2m_N}(\beta)$ or $y_k = y_{2m_N+1}(\beta)$, most of the measurement outcomes, except for measuring outcomes with a small number of photons including vacuum, can generate SCSs of certain amplitude with slightly different fidelities which can reduce the requirements for the PNR detector to resolve the number of photons. The idea of generating even/odd SCSs of large amplitude for almost any measurement outcome $n$ greater than a certain value $n_0$, i.e. $n > n_0$ is promising and deserves separate consideration.

The success probabilities to conditionally generate $2m_1, 2m_1 + 1$ CV states in Eqs. (23,24) are given by

$$P_{2m_1}^{(0)} = \frac{1}{\cosh s}\left(\frac{1-t_1^2}{t_1^2}\right)^{2m_1} \frac{y_1^{2m_1}}{(2m_1)!} Z^{(2m_1)}(y_1), \qquad (4)$$

$$P_{2m_1+1}^{(0)} = \frac{1}{\cosh s}\left(\frac{1-t_1^2}{t_1^2}\right)^{2m_1+1}\frac{y_1^{2m_1+1}}{(2m_1+1)!}Z^{(2m_1+1)}(y_1), \qquad (5)$$

in the case of using one BS and PNR detector. It is worth noting that the success probability of the measurement also depends on the BS parameter $t_1$, in contrast to the fidelity, which is completely determined by the value $y_1$. To evaluate the success probability to implement the even/odd SCSs one should take the value $y_1$ either $y_1 = y_{2m_N}(\beta)$ in Eq. (4) to get $P_{2m_1}^{(0)}\left(y_{2m_N}(\beta)\right)$ or $y_1 = y_{2m_N+1}(\beta)$ in Eq. (5) to obtain $P_{2m_1+1}^{(0)}\left(y_{2m_N+1}(\beta)\right)$. The value $1/\cosh s$ must be chosen taking into account the relationship between the values $y_0$ and $y_1$. The success probabilities both for arbitrary values of $y_1$ and for those providing SCSs generation fall rather quickly with increasing the parameter $m_1$, i.e. $P_{2m_1+2}^{(0)}/P_{2m_1}^{(0)} = ((1-t_1^2)/t_1^2)^2 y_1^2 ((2m_1)!/(2m_1+2)!)\left(Z^{(2m_1+2)}(y_1)/Z^{(2m_1)}(y_1)\right) \ll 1$,
$P_{2m_1+3}^{(0)}/P_{2m_1+1}^{(0)} = ((1-t_1^2)/t_1^2)^2 y_1^2((2m_1+1)!/(2m_1+3)!)$
$\left(Z^{(2m_1+3)}(y_1)/Z^{(2m_1+1)}(y_1)\right) \ll 1$ involving the case
$y_1 = y_{2m_N}(\beta)$ and $y_1 = y_{2m_N+1}(\beta)$. The finding drastically reduces the generation rate of the even/odd SCSs of larger amplitude requiring more photon subtraction in setup with one BS and PNR detector.

Let us compare the success probabilities $P_{n_1,n_2,\ldots,n_k}^{(0,0,\ldots,0)}$ of the even/odd SCSs generation in configuration with $k$ BSs and PNR detectors and ones $P_{n_1+n_2+\cdots+n_k}^{(0)}$ by equal subtraction of photons in both cases. Here, a superscript $(0)$ indicates on the hub with one BS and PNR detector while a superscript $(0,0,\ldots,0)$ with $k$ zeros is applied to show that $k$ BSs and PNR detectors are used in the optical scheme in Fig. 1. The success probability to implement the CV states of definite parity in Eqs. (23,24) becomes

$$P_{n_1,n_2,\ldots,n_k}^{(0,0,\ldots,0)} = \frac{1}{\cosh s}\prod_{l=1}^{k}\left(\frac{1-t_l^2}{t_l^2}\right)^{n_l}\frac{y_l^{n_l}}{n_l!}Z^{(n_1+n_2+\cdots+n_k)}(y_k) =$$
$$P_{n_k/n_1,n_2,\ldots,n_{k-1}}^{(0)}P_{n_{k-1}/n_1,n_2,\ldots,n_{k-2}}^{(0)}\cdots P_{n_3/n_1,n_2}^{(0)}P_{n_2/n_1}^{(0)}P_{n_1}^{(0)}, \qquad (6)$$

where the conditional probability is given by

$$P_{n_i/n_1,n_2,\ldots,n_{i-1}}^{(0)} = \left(\frac{1-t_i^2}{t_i^2}\right)^{n_i}\frac{y_i^{n_i}}{(n_i)!}\frac{Z^{(n_1+n_2+\cdots+n_i)}(y_i)}{Z^{(n_1+n_2+\cdots+n_{i-1})}(y_{i-1})}, \qquad (7)$$

obeying the normalization condition $\sum_{n_i=0}^{\infty}P_{n_i/n_1,n_2,\ldots,n_{i-1}}^{(0)} = 1$. Here, the quantity $P_{n_1}^{(0)}$ is presented in Eqs. (4,5). The success probabilities for generating even/odd SCSs is obtained by taking either $y_k = y_{2m_N}(\beta)$ or $y_k = y_{2m_N+1}(\beta)$ heeding the relationship between $y_i$ and $y_k$ ($i<k$) in Eq. (6) to get amounts $P_{n_1,n_2,\ldots,n_k}^{(0,0,\ldots,0)}\left(y_{2m_N}(\beta)\right)$ and $P_{n_1,n_2,\ldots,n_k}^{(0,0,\ldots,0)}\left(y_{2m_N+1}(\beta)\right)$, respectively. Substituting either $y_{2m_N}(\beta)$ or $y_{2m_N+1}(\beta)$ instead of $y_k$, one estimates the ratio of the success probabilities for the hub with one element $P_{n_1+n_2+\cdots+n_k}^{(0)}\left(y_{2m_N,2m_N+1}(\beta)\right)$ and $k$ elements $P_{n_1,n_2,\ldots,n_k}^{(0,0,\ldots,0)}\left(y_{2m_N,2m_N+1}(\beta)\right)$ configured to generate even/odd SCSs

$$\frac{P_{n_1,n_2,\ldots,n_k}^{(0,0,\ldots,0)}\left(y_{2m_N,2m_N+1}(\beta)\right)}{P_{n_1+n_2+\cdots+n_k}^{(0)}\left(y_{2m_N,2m_N+1}(\beta)\right)} = (t^{-2})^{(k-1)n_1+(k-2)n_2+\cdots+n_{k-1}}\frac{(n_1+n_2+n_3\ldots+n_k)!}{(n_1)!(n_2)!\ldots(n_k)!}, \qquad (8)$$

with the same number of extracted photons in both cases, where the hub with $k$ elements composes of identical BSs $t_1 = t_2 \ldots = t_k = t$. When evaluating ratio in Eq. (12), we supposed that $y_0^{(1)} = y_0^{(k)}$ which is acceptable to compare two values. As can be seen from the relation, the following estimate $P_{n_1,n_2,\ldots,n_k}^{(0,0,\ldots,0)}\left(y_{2m_N,2m_N+1}(\beta)\right) \gg P_{n_1+n_2+\cdots+n_k}^{(0)}\left(y_{2m_N,2m_N+1}(\beta)\right)$ takes place, since there are two factors postulating the inequality. The factor $(n_1+n_2+n_3\ldots+n_k)!/(n_1)!(n_2)!\ldots(n_k)!$

makes a significant increasing contribution to the ratio of the probabilities in addition to the increasing multiplier $(t^{-2})^{(k-1)n_1+(k-2)n_2+\cdots+n_{k-1}}$. But it is worth heeding the decrease in the transmittance coefficient $t$ is limited by the corresponding condition on amount $y_0^{(k)}$ noted above. Substantial reducing the transmittance coefficient $t$ can lead to impossibility of generating even/odd SCSs of certain amplitude with required high fidelity. In general, the constraint on initial amount $y_0^{(k)}$ can significantly limit the possibility of the probability-efficient SCSs generation. Note also the inequality under study can be only increased in the case of $t_1 > t_2 > \cdots t_k$.

Let us consider in more detail the success probability to implement SCSs in a scheme with two identical BSs ($t_1 = t_2 = t$) and two PNR detectors. Then, for example, the success probability to generate even SCS is given by

$$P_{2m_1,2m_2}^{(0,0)} = \sqrt{1 - 4\left(\frac{y_{2m_N}(\beta)}{t^4}\right)^2 \left(\frac{1-t^2}{t^2}\right)^{2m_1+2m_2}}$$
$$\frac{1}{(t^2)^{2m_1}} \frac{\left(y_{2m_N}(\beta)\right)^{2m_1+2m_2}}{(2m_1)!(2m_2)!} Z^{(2m_1+2m_2)}\left(y_{2m_N}(\beta)\right), \qquad (9)$$

in the case of total even number of detected photons $2m_1 + 2m_2 = 2m_N$ provided that both detectors have registered even Fock states $2m_1$ and $2m_2$, respectively. Here, the multiplier $1/\cosh s$ is expressed in terms of the parameter $y_2 = y_{2m_N}(\beta)$, i.e. $y_2 = t^4 y_0 = y_{2m_N}(\beta)$. In figures 4(a,b), we show the dependence of the success probabilities to generate even SCSs for various even measurement outcomes registered by two PNR detectors provided that the total number of registered photons is 20, i.e. $2m_1 + 2m_2 = 20$. Here, the SCS amplitude range is selected from 2.2 to 3, where the appropriate large fidelity (Fig. 2(a)) is provided. Note that a further increase of the SCS amplitude $\beta > 3$ may lead to the impossibility of generating the corresponding SCS with the fidelity shown in Fig. 2(a) for the selected values of the BSs parameters $t_1 = t_2 = t$ since the value $y_2/t^4$ may exceed the threshold value 0.5. Numerical results confirm significant gain in the success probability when using two PNR detectors compared to one. All probabilities $P_{n_1,n_2}^{(0,0)}$ in Figs. 4(a,b) with $n_1 \neq 0$ exceed the probability to implement even SCS with one PNR detector. The dependence of $P_{0,2m_N}^{(0,0)}\left(y_{2m_N}(\beta)\right) \cong P_{2m_N}^{(0)}\left(y_{2m_N}(\beta)\right)$ (the difference between the probabilities is very insignificant and is related to the different values of input $y_0^{(2)} > y_0^{(1)}$ that should be applied that entails only slight increase of $P_{2m_N}^{(0)}\left(y_{2m_N}(\beta)\right)$ over $P_{0,2m_N}^{(0,0)}\left(y_{2m_N}(\beta)\right)$, i.e. $P_{2m_N}^{(0)}\left(y_{2m_N}(\beta)\right) > P_{0,2m_N}^{(0,0)}\left(y_{2m_N}(\beta)\right)$) is shown in Fig. 4(a,b) and takes on the lowest possible values of all the plots presented. As can be seen from the plots, the probability $P_{10,10}^{(0,0)}$ significantly exceeds all other probabilities $P_{n_1,n_2}^{(0,0)}$ with $n_1 \neq n_2$ for definite values of $\beta$. The use of the BSs with a little lower transmittance coefficient (Fig. 4(b)) makes it possible to increase the success probability $P_{10,10}^{(0,0)}$ by almost two orders compared $P_{10,10}^{(0,0)}$ in Fig. 4(a). Note also that the fidelity of the output state solely depends on the total number of extracted photons either $2m_N$ or $2m_N + 1$, in contrast to the success probability, which is also determined by the number of photons extracted by each PNR detector.

Thus, use of the multiple hub can provide, at least, two advantages over optical scheme with a single BS and PNR detection. PNR detectors are characterized by several different parameters, such as efficiency, maximum number of the detectable photons, photon number resolution, etc. Registration of a large number of photons by a single PNR detector can impose too high requirements on maximum number of detectable photons detected with single-photon resolution which can be quite difficult to implement in practice. The multiphoton state splitting realized by multi-element hub through a split of the measured light into several output measurement modes each of which measures a smaller number of photons avoids higher requirements for the PNR detection. So, the multiphoton state demultiplexing with two PNR detectors registering up to 10 photons with single-photon resolution each is more feasible compared to one PNR detector, which should resolve up to 20 photons with the same resolution. An even more advantageous situation can take place in the case of exact detection maximum, say, up to 50 photons by one PNR detector. Instead of measuring 50 photons by one PNR detector, one can make use of, for example, a hub with 5 BSs and PNR detectors capable of resolving up to 10 photons with single-photons resolution each. In addition to reducing the sensitivity requirements for the PNR detectors in a scheme with a large number of elements, multi-element hub provides a significant gain in the success probability. The gain can only exponentially increase with an increment in the number of elements in the hub.

The discussion presented above referred to the ideal operation of the measuring technique. Therefore, the obtained numerical results in the figures (2-4) can be classified as perfect, that is, those that can be observed under ideal experimental conditions without taking into account the imperfection of the experimental measurement technique. Taking into account the imperfection of measuring equipment can worsen the values of perfect parameters (SCS amplitude $\beta$, fidelity). Nevertheless, consideration of the optical scheme in Figure 1 under ideal conditions makes sense, since it allows one to find the perfect values of the parameters that one should strive for in the practical implementation of quantum engineering of even/odd SCSs.

**Influence of the quantum efficiency of the PNR detector on perfect values of the SCS generator.** In real experiments, PNR detectors can have rather high quantum efficiency, but not an ideal one, i.e. $\eta < 1$ that can lead to a deterioration of the characteristics of the output states. To estimate the level of the fidelity degradation of generated SCSs, one introduces even/odd positive-operator values measure (POVM) elements of the PNR detector

$$\Pi_{2m} = \sum_{x=0}^{\infty} \begin{pmatrix} C_{2(m+x)}^{2m} \eta^{2m}(1-\eta)^{2x}|2(m+x)\rangle\langle 2(m+x)| + \\ C_{2(m+x)+1}^{2m} \eta^{2m}(1-\eta)^{2x+1}|2(m+x)+1\rangle\langle 2(m+x)+1| \end{pmatrix}, \quad (10)$$

$$\Pi_{2m+1} = \sum_{x=0}^{\infty} \begin{pmatrix} C_{2(m+x)+1}^{2m+1} \eta^{2m+1}(1-\eta)^{2x}|2(m+x)+1\rangle\langle 2(m+x)+1| + \\ C_{2(m+x+1)}^{2m+1} \eta^{2m+1}(1-\eta)^{2x+1}|2(m+x+1)\rangle\langle 2(m+x+1)| \end{pmatrix}, \quad (11)$$

where $C_i^j$ is a binomial coefficient. Now, the fidelities $Fid_{2m}^{(0)}(\eta) = tr\left(\rho_{2m}^{(0)}(|SCS_+\rangle\langle SCS_+|)\right)$ and $Fid_{2m+1}^{(0)}(\eta) = tr\left(\rho_{2m+1}^{(0)}(|SCS_-\rangle\langle SCS_-|)\right)$, where $tr$ means the trace operation, $\rho_{2m}^{(0)} = tr_2\left(\rho^{(0)}\Pi_{2m}(\eta)\right)$, $\rho_{2m+1}^{(0)} = tr_2\left(\rho^{(0)}\Pi_{2m+1}(\eta)\right)$ are the conditional states and $\rho^{(0)} = \left(BS_{12}(|SMSV\rangle_1|0\rangle_2)\right)\left(BS_{12}(|SMSV\rangle_1|0\rangle_2)\right)^+$ is an original one, where upper symbol + in operator is responsible for the Hermitian conjugation operation, should be analyzed by their decomposing in terms of the powers of small parameter $1-\eta$ up to $(1-\eta)^2$, provided that quantum efficiency of modern PNR detectors is high enough $\eta \approx 1$ but $\eta < 1$

$$Fid_{2m}^{(0)}(\eta) = Fid_{2m}^{(0)}(\eta=1)\left(1-(1-\eta)\frac{1-t_1^2}{t_1^2}\langle n\rangle_{2m} + (1-\eta)^2 f_{2,2m}^{(0)}\right), \quad (12)$$

$$Fid_{2m+1}^{(0)}(\eta) = Fid_{2m+1}^{(0)}(\eta=1)\left(1-(1-\eta)\frac{1-t_1^2}{t_1^2}\langle n\rangle_{2m+1} + (1-\eta)^2 f_{2,2m+1}^{(0)}\right), \quad (13)$$

where the mean number of photons of the even/odd SCSs

$$\langle n\rangle_{2m} = y\frac{Z^{(2m+1)}}{Z^{(2m)}}, \quad (14)$$

$$\langle n \rangle_{2m+1} = y \frac{Z^{(2m+2)}}{Z^{(2m+1)}}, \qquad (15)$$

is calculated with $y = y_{2m_N}(\beta)$ for even SCSs and $y = y_{2m_N+1}(\beta)$ for odd SCSs. Second order terms in decompositions in Eqs. (12,13) are defined as

$$f^{(0)}_{2,2m} = \frac{\langle n \rangle_{2m}}{2}\left(\frac{1-t_1^2}{t_1^2}\right)^2 \left(2\langle n \rangle_{2m} - \langle n \rangle_{2m+1}\left(1 - R^{(0)}_{2m}\right)\right), \qquad (16)$$

$$f^{(0)}_{2,2m+1} = \frac{\langle n \rangle_{2m+1}}{2}\left(\frac{1-t_1^2}{t_1^2}\right)^2 \left(2\langle n \rangle_{2m+1} - \langle n \rangle_{2m+2}\left(1 - R^{(0)}_{2m+1}\right)\right), \qquad (17)$$

where $R^{(0)}_{2m} = \left|\langle SCS_+|\Psi^{(0)}_{2m+2}(y_{2m})\rangle\right|^2 / \left|\langle SCS_+|\Psi^{(0)}_{2m}(y_{2m}(\beta))\rangle\right|^2$ and $R^{(0)}_{2m+1} = \left|\langle SCS_-|\Psi^{(0)}_{2m+3}(y_{2m+1}(\beta))\rangle\right|^2 / \left|\langle SCS_-|\Psi^{(0)}_{2m+1}(y_{2m+1}(\beta))\rangle\right|^2$. Here, the fidelities $Fid^{(0)}_{2m}(\eta = 1)$ and $Fid^{(0)}_{2m+1}(\eta = 1)$ are those that are shown in Figs. 2. The success probabilities in detecting $2m, 2m+1$ photons and thereby generating even/odd SCSs in the output mode are given by

$$P^{(0)}_{2m}(\eta) = \frac{\eta^{2m}}{\cosh s}\left(\frac{1-t_1^2}{t_1^2}\right)^{2m} \frac{y_1^{2m}}{(2m)!} \left(\begin{array}{c} \sum_{x_1=0}^{\infty}\left(\frac{1-t_1^2}{t_1^2}\right)^{2x_1}\frac{y_1^{2x_1}}{(2x_1)!} Z^{(2(m+x_1))}(y_1)(1-\eta)^{2x_1} + \\ \sum_{x_1=0}^{\infty}\left(\frac{1-t_1^2}{t_1^2}\right)^{2x_1}\frac{y_1^{2x_1}}{(2x_1)!} Z^{(2(m+x_1))}(y_1)(1-\eta)^{2x_1} \end{array}\right) \approx$$

$$\eta^{2m} P^{(0)}_{2m}(\eta = 1)\left(1 + (1-\eta)\frac{1-t_1^2}{t_1^2}\langle n\rangle_{2m}\right), \qquad (18)$$

$$P^{(0)}_{2m+1}(\eta) = \frac{\eta^{2m+1}}{\cosh s}\left(\frac{1-t_1^2}{t_1^2}\right)^{2m+1} \frac{y_1^{2m+1}}{(2m+1)!}$$

$$\left(\begin{array}{c} \sum_{x_1=0}^{\infty}\left(\frac{1-t_1^2}{t_1^2}\right)^{2x_1}\frac{y_1^{2x_1}}{(2x_1)!} Z^{(2(m+x_1)+1)}(y_1)(1-\eta)^{2x_1} + \\ \sum_{x_1=0}^{\infty}\left(\frac{1-t_1^2}{t_1^2}\right)^{2x_1+1}\frac{y_1^{2x_1+1}}{(2x_1+1)!} Z^{(2(m+x_1+1))}(y_1)(1-\eta)^{2x_1+1} \end{array}\right) \approx$$

$$\eta^{2m+1} P^{(0)}_{2m+1}(\eta = 1)\left(1 + (1-\eta)\frac{1-t_1^2}{t_1^2}\langle n\rangle_{2m+1}\right), \qquad (19)$$

where $P^{(0)}_{2m}(\eta = 1)$ and $P^{(0)}_{2m+1}(\eta = 1)$ are given by Eqs. (4,5). As can be seen from the presented formulas, the fidelity is a deviation by a certain amount from the perfect value $\left(F^{(0)}_{2m,2m+1}(\eta = 1) > F^{(0)}_{2m,2m+1}(\eta)\right)$ in the case of small values of the quantum inefficiency $1 - \eta \ll 1$. The success probability is increased by the same amount compared to the perfect value $P^{(0)}_{2m,2m+1}(\eta) > P^{(0)}_{2m,2m+1}(\eta = 1)$.

The largest decreasing contribution to the fidelities in Eqs. (12,13) proportional to $1 - \eta$ includes the product of the BS parameter $(1 - t_1^2)/t_1^2$ by the mean number of photons either $\langle n \rangle_{2m}$ or $\langle n \rangle_{2m+1}$. The mean number of photons in SCSs is known to be proportional to its amplitude squared i.e. $\langle n \rangle_{SCS} \sim \beta^2$ and the more the SCS amplitude is, the greater the average number of photons is in it. To check the fact for the CV states of certain parity, the dependences of the mean number of photons in the generated states are shown in the figure 5. The plots in Figure 5(a,b) are the dependence of the mean number of photons on the parameter $y_{2m_N}(\beta), y_{2m_N+1}(\beta)$ which provides the fidelities in Fig. 2, while the graphs in Figure 5(c,d) also show the mean number of photons in the CV states of definite parity but in dependence on the SCS amplitude $\beta$. The form of the curves in Figure 5(a,b) is related to the fact that their arguments are defined in different ranges of change $y$ (see Figs. 3(a,b)). So, the arguments $y_{90}(\beta)$ and $y_{91}(\beta)$ for the states $|\Psi^{(0,0,\ldots,0)}_{90}\rangle$ and $|\Psi^{(0,0,\ldots,0)}_{91}\rangle$, respectively, change in a smallest range (Figs. 3(a,b)), which entails that the plots of $\langle n \rangle_{90}$ and $\langle n \rangle_{91}$ are maximally shifted to the left. As can be seen the Figs. 5(c,d) the condition $\langle n \rangle_{2m} \approx \beta^2, \langle n \rangle_{2m+1} \approx \beta^2$ is met (the curves on the graphs almost coincide regardless of the number of extracted photons), which also indicates the that the generated CV states of a certain parity can approximate even/odd SCSs with high fidelity (Figs. 2). The maximum number of average number of photons is limited to a value of just over 35 for the maximum considered number of subtracted photons, namely 90 and 91. Note that the graphs in Figure 5 are applicable to a hub regardless of the number of its elements, that is, both to the hub with one and $k$ elements.

The contribution of the mean number of photons can be partly compensated for by the BS parameter provided that a beam splitter different from the balanced one that is, with $t_1 > 1/\sqrt{2}$ is used to guarantee the performance of the condition $(1 - t_1^2)/t_1^2 < 1$. Then, the factor reducing the fidelity can be reduced to zero in the case of use of the HTBS with $t_1 \to 1$, when the BS parameter $(1 - t_1^2)/t_1^2$ tends to zero, thereby nullifying contribution of the mean number of photons to the overall fidelity in Eqs. (12,13) that can provide perfect values of the fidelity, i.e. $Fid_{2m}^{(0)}(\eta) \approx Fid_{2m}^{(0)}(\eta = 1)$ and $Fid_{2m+1}^{(0)}(\eta) \approx Fid_{2m+1}^{(0)}(\eta = 1)$. Unfortunately, the strategy of using HTBS has a significant drawback. It follows from the formulas (18,19) that the success probability just depends on the same BS multiplier $(1 - t_1^2)/t_1^2$ in the appropriate degree either $2m$ or $2m + 1$ as well as its perfect values in Eqs. (4,5). The power dependence can very quickly reduce the perfect values of the success rate and to zero which confirms the inapplicability of the HTBS strategy to extracting a large number of photons from the initial SMSV state. Indeed, the HTBS redirects a single photon into measurement mode with probability $\sim(1 - t_1^2)$, while $n$ photons can be redirected with probability $\sim(1 - t_1^2)^n \approx 0$ in the case of $t_1 \to 1$. In general, the relationship between the fidelity of the output CV state and its success probability can be reflected by the following expressions:
$\Delta F_{2m}^{(0)}(\eta) \cdot \Delta P_{2m}^{(0)}(\eta) \approx (1 - \eta)^2((1 - t_1^2)/t_1^2)^2 \langle n \rangle_{2m} F_{2m}^{(0)}(\eta = 1) \cdot P_{2m}^{(0)}(\eta = 1)$ and
$\Delta F_{2m+1}^{(0)}(\eta) \cdot \Delta P_{2m+1}^{(0)}(\eta) \approx (1 - \eta)^2((1 - t_1^2)/t_1^2)^2 \langle n \rangle_{2m+1} F_{2m+1}^{(0)}(\eta = 1) \cdot P_{2m+1}^{(0)}(\eta = 1)$,
where $\Delta F_{2m,2m+1}^{(0)}(\eta) = F_{2m,2m+1}^{(0)}(\eta = 1) - F_{2m,2m+1}^{(0)}(\eta)$ and $\Delta P_{2m,2m+1}^{(0)}(\eta) = P_{2m,2m+1}^{(0)}(\eta) - P_{2m,2m+1}^{(0)}(\eta = 1)$ are the differences between the perfect and those values that are realized in the case of using an inefficient PNR detector.

The result can be extended to the case of an optical hub consisting of $k$ elements, which leads to rather cumbersome expressions for the fidelity and the success probability. Nevertheless, for qualitative conclusions, it is sufficient to consider the case of $k$ detectors with the same quantum efficiency $\eta$ and limit our consideration to the first order in quantum inefficiency $1 - \eta$, which gives

$$Fid_{2m_N,2m_N+1}^{(0,0,\dots,0)}(\eta) = Fid_{2m_N,2m_N+1}^{(0,0,\dots,0)}(\eta = 1)\left(1 - (1 - \eta)\frac{1 - t_1^2 t_2^2 \dots t_k^2}{t_1^2 t_2^2 \dots t_k^2} \langle n \rangle_{2m_N,2m_N+1}\right), \quad (20)$$

$$P_{2m_N,2m_N+1}^{(0,0,\dots,0)}(\eta) = P_{2m_N,2m_N+1}^{(0,0,\dots,0)}(\eta = 1)\left(1 + (1 - \eta)\frac{1 - t_1^2 t_2^2 \dots t_k^2}{t_1^2 t_2^2 \dots t_k^2} \langle n \rangle_{2m_N,2m_N+1}\right), \quad (21)$$

where $Fid_{2m_N,2m_N+1}^{(0,0,\dots,0)}(\eta = 1)$ and $P_{2m_N,2m_N+1}^{(0,0,\dots,0)}(\eta = 1)$ are the perfect values of the fidelity and the success probability. It follows from the expressions that the values of the fidelity and success probability act in discord with respect to each other, that is, the fidelity decreases and the success probability increases by the same amount. The value is proportional both to the mean number of photons in the generated state, which follow from the curves in Figure 5, and to the hub's parameter $(1 - t_1^2 t_2^2 \dots t_k^2)/t_1^2 t_2^2 \dots t_k^2$. In general, the multiplier $(1 - t_1^2 t_2^2 \dots t_k^2)/t_1^2 t_2^2 \dots t_k^2$ is more than the factor $(1 - t_1^2)/t_1^2$ present in Eqs. (16,17,22,23), i.e. $(1 - t_1^2 t_2^2 \dots t_k^2)/t_1^2 t_2^2 \dots t_k^2 \geq (1 - t_1^2)/t_1^2$ which reduces the possibilities for the hub parameter to decrease the contribution of the mean number of photons. So, in the case of using the same BSs so that the condition $t_1 = t_2 = \dots = t_k = t$ takes place, the hub's parameter becomes lowering $((1 - t^{2k})/t^{2k} < 1)$ in the case of $t > 1/\sqrt[2k]{2}$. As in the case of a one-element hub, it is possible to achieve perfect values for the fidelity $Fid_{2m_N,2m_N+1}^{(0,0,\ldots,0)}(\eta) \approx Fid_{2m_N,2m_N+1}^{(0,0,\ldots,0)}(\eta = 1)$ in the case of $(1 - t_1^2 t_2^2 \ldots t_k^2)/t_1^2 t_2^2 \ldots t_k^2 \to 1$ which is guaranteed to lead to a strategy of using $k$ HTBSs with $t_1 \to 1, t_2 \to 1, \ldots, t_k \to 1$. The strategy with HTBSs is hardly practical for the quantum state engineering of the even/odd SCSs, since the perfect values of the success probability tend to zero in the case of subtracting a large number of photons (say, more of 20) even though (as shown above) the strategy with multiphoton state demultiplexing gives a certain gain to the final success probability (Eq. (21)) due to the multiplier $(1 - \eta)((1 - t_1^2 t_2^2 \ldots t_k^2)/t_1^2 t_2^2 \ldots t_k^2)\langle n \rangle_{2m_N,2m_N+1}$. However, the strategy with HTBSs (or partial use of HTBSs) may become practical when extracting a small number of photons (say, less than 10). It is also possible to estimate deviations of the parameter values from their perfect values taking into account $Fid_{2m,2m+1}^{(0)}(\eta = 1) > Fid_{2m,2m+1}^{(0)}(\eta)$ and $P_{2m,2m+1}^{(0)}(\eta) > P_{2m,2m+1}^{(0)}(\eta = 1)$, namely, $\Delta F_{2m_N}^{(0,0,\ldots,0)}(\eta) \cdot \Delta P_{2m_N}^{(0,0,\ldots,0)}(\eta) \approx (1 - \eta)^2((1 - t_1^2 t_2^2 \ldots t_k^2)/t_1^2 t_2^2 \ldots t_k^2)^2 \langle n \rangle_{2m_N} F_{2m_N}^{(0,0,\ldots,0)}(\eta = 1) \cdot P_{2m_N}^{(0,0,\ldots,0)}(\eta = 1)$ and $\Delta F_{2m_N+1}^{(0,0,\ldots,0)}(\eta) \cdot \Delta P_{2m_N+1}^{(0,0,\ldots,0)}(\eta) \approx (1 - \eta)^2((1 - t_1^2 t_2^2 \ldots t_k^2)/t_1^2 t_2^2 \ldots t_k^2)^2 \langle n \rangle_{2m_N+1} F_{2m_N+1}^{(0,0,\ldots,0)}(\eta = 1) \cdot P_{2m_N+1}^{(0,0,\ldots,0)}(\eta = 1)$, where the quantities $\Delta F_{2m_N,2m_N+1}^{(0,0,\ldots,0)}(\eta)$ and $\Delta P_{2m_N,2m_N+1}^{(0,0,\ldots,0)}(\eta)$ are the differences between the perfect and those values that are obtained in the practical case of using an inefficient PNR detector $\Delta F_{2m_N,2m_N+1}^{(0,0,\ldots,0)}(\eta) = F_{2m_N,2m_N+1}^{(0,0,\ldots,0)}(\eta = 1) - F_{2m_N,2m_N+1}^{(0,0,\ldots,0)}(\eta)$ and $\Delta P_{2m_N,2m_N+1}^{(0,0,\ldots,0)}(\eta) = P_{2m_N,2m_N+1}^{(0,0,\ldots,0)}(\eta) - P_{2m_N,2m_N+1}^{(0,0,\ldots,0)}(\eta = 1)$, respectively.

The dependences in Figure 5 allow us to estimate the contribution of the decreasing term to the fidelity of the output states generated with inefficient PNR detectors. Let us take the maximum observed value $\langle n \rangle_{90} = \langle n \rangle_{91} \approx 35$ for estimation. Then, the decreasing factor can be evaluated by $\approx 35((1 - t_1^2)/t_1^2)$ in a scheme with one BS and PNR detector (Eqs. (12,13,18,19)) and $\approx 35((1 - t_1^2 t_2^2)/t_1^2 t_2^2)$ (Eqs. (20,21)) in setup with two BSs and PNR detectors in the case subtraction of either 90 for generation of even SCS or 91 photons for generation of odd SCS. The reducing factor $35((1 - t_1^2)/t_1^2)$ can take the following values: 8.21 for $t_1 = 0.9$; 3.78 for $t_1 = 0.95$ and 1.44 for $t_1 = 0.98$. In the case of a hub with two identical BSs ($t_1 = t_2 = t$) and PNR detectors, the decreasing multiplier $35((1 - t_1^2 t_2^2)/t_1^2 t_2^2)$ can take on the following values: 18.35 for $t = 0.9$; 7.97 for $t = 0.95$ and 2.94 for $t = 0.98$. If we take value of the quantum efficiency $\eta = 0.98^{33}$, then one have the following fidelities: $Fid_{90,91}^{(0)}(\eta) \approx 0.8358 \cdot Fid_{90,91}^{(0)}(\eta = 1)$ for $t_1 = 0.9$; $Fid_{90,91}^{(0)}(\eta) \approx 0.9244 \cdot Fid_{90,91}^{(0)}(\eta = 1)$ for $t_1 = 0.95$; $Fid_{90,91}^{(0)}(\eta) \approx 0.9712 \cdot Fid_{90,91}^{(0)}(\eta = 1)$ for $t_1 = 0.98$. One can choose the value of the SCS amplitude $\beta$ in such a way that $Fid_{90,91}^{(0)}(\eta = 1) > 0.99$ (Figs. 2), but it does not corroborate the utility of the strategy, since the success probability is proportional to either $(1 - t_1^2)^{90}$ or $(1 - t_1^2)^{91}$ and can take on very small values, much less of $< 10^{-20}$. Comparing the values of decreasing multipliers, it can be noted that they can only increase with an increase in the number of the hub's elements. So, one can evaluate from Eqs. (20,21) $Fid_{90}^{(0,0)}(\eta) \approx 0.8406 \cdot Fid_{45,45}^{(0,0)}(\eta = 1)$ for $t_1 = t_2 = 0.95$ and $\eta = 0.98$.

In order to take advantage of two-element hubs in terms of substantial gain in the success probability keeping the fidelity of the output SCS at an acceptable level, it is worth reducing the number of photons subtracted. As follows from Figure 4, an increase in the success probability is provided when two PNR detectors detect the same or almost the same number of photons. So, in the case of generating even SCS of amplitude of $\beta = 3$ by detecting 20 photons with two PNR detectors, we approximately have $\langle n \rangle_{20} \approx 8$ and $Fid_{20}^{(0,0)}(\eta = 0.98) \approx 0.9162 \cdot Fid_{10,10}^{(0,0)}(\eta = 1)$ for $t_1 = t_2 = 0.9$. The success probability of the event takes the values $\sim 10^{-9}$. If one reduces the number of extracted photons by half (say 10 photons) with help of two BSs with $t_1 = t_2 = 0.95$, then one can estimate the final reduction factor as 0.9727 and the output fidelity becomes $Fid_{20}^{(0,0)}(\eta = 0.98) \approx 0.9727 \cdot Fid_{5,5}^{(0,0)}(\eta = 1)$. The success probability of the event is estimated at the level of $\sim 10^{-7}$ which can be evaluated as more practical in the quantum engineering of even/odd SCSs of amplitudes $\beta = 2.5$. Progress in the development of the qualitative PNR detectors with quantum efficiency $\eta > 0.98$ can improve the above estimates. Note that the term acceptable is used in comparison with the success probability of spontaneous parametric down conversion (SPDC) whose effectiveness is usually bounded above by a value $10^{-6}$. In a real experimental case, the efficiency of the SPDC can be even lower, on the order of $10^{-8}$.

**Discussion** Study of quantum effects in physical systems of macroscopic sizes is largely driven by introduction of Schrödinger cat states[1] as one of the most fundamental issue of quantum mechanics. It is no occasional that solution of the problem spurred serious interest to multiphoton states that could contain, on average, larger number of photons not limiting to a few to observe nonclassical features on the macroscopic states. Here, we have demonstrated the possibility of generating a whole family of multiphoton states of a certain parity generated from related to them SMSV state by extraction of multiphoton state from original state and registering them in the measurement modes of a multi-element hub. Redirecting photons in an indistinguishable manner, followed by probabilistic detection of a certain number of photons, redistributes the input distribution of the SMSV state to a new (Eqs. (23,24)) associated with the original. New CV states of definite parity with a larger mean number of photons have potential applications in quantum state engineering and optical quantum metrology. The generated states of the CV family of definite parity are characterized by only one parameter $0 < y < 0.5$, which exclusively depends on the squeezing amplitude of the original SMSV. In the scheme with one BS and PNR detector, the input parameter $y_0$ is multiplied by the BS transmittance coefficient squared, thereby lowering it by the corresponding value. In the scheme with multiphoton state demultiplexing, the reduction of $y_0$ can be more destructive due to successive multiplication by BS transmittance coefficients squared. We have shown that for each multiphoton state $|m_N\rangle$ subtracted from original SMSV there is a parameter value $y_{2m_N}(\beta)$ for even and $y_{2m_N+1}(\beta)$ for odd number of subtracted photons which provides the maximum fidelity of the CV state of a certain parity with even/odd SCS of amplitude $\beta$. In order to achieve the perfect fidelity of the SCSs generator, it is necessary to ensure the fulfillment of the condition $y_k = y_{2m_N,2m_N+1}(\beta)$. Thus, the squeezing amplitude $s$ and BS's parameters $t_i$ can be chosen appropriately to ensure the condition $y_k = y_{2m_N,2m_N+1}(\beta)$. From a practical point of view, the important point is that the values of $y_{2m_N,2m_N+1}(\beta)$ decrease with an increase of the number of the detected photons.

Control over the success probability is possible since it largely depends on the number of photons extracted by each detector, in contrast to the fidelity of the output state, which is solely determined by the total number of photons detected. Therefore, use of the multiphoton state demultiplexing by means of use of multi-element hub makes sense to increase the success probability without affecting the fidelity of the output state. The use of only one BS and detector leads to a rather small values of the success probability $(10^{-20} >)$ of even/odd SCSs generation of large amplitude, when a sufficiently large number $(> 50)$ of photons is subtracted. Greater redirection of photons of original SMSV state into measurement modes due to increased reflectance coefficients of the BSs is additional factor raising the success probability of the required event. But the strategy with BSs with increased reflectance coefficients should be accompanied by an increase in the squeezing amplitude of the initial SMSV and has a restrictive effect due to the inability to achieve the required condition $y_k = y_{2m_N,2m_N+1}(\beta)$ if a large number of BSs is used. In addition, use of several PNR detectors instead of one reduces the requirements imposed on the maximum number of detected photons registered with single-photon resolution. In general, strategy with multiple hub is effective and promising for the quantum engineering of high-fidelity≥ 0.99 even/odd SCSs of amplitude≥ 5 with perfect PNR detectors. Note the number of the subtracted photons in the SCSs generator with ideal PNR detectors can be increased (say up to $10^6$) which leads to an increase in the amplitude of the generated even/odd SCSs.

In the case of practical quantum engineering of even/odd SCSs with inefficient PNR detectors, fidelity and success probability can become competing parameters which is analytically expressed. Use of the highly transmitting BSs generates the even/odd SCSs with fidelity close to perfect but only at the expense of a sharp decrease of the success probability since the multiphoton state has less chance of appearing in the measurement modes. An increase in the success probability is possible by redirecting more photons into the measurement modes but it reduces the fidelity of the even/odd SCSs. A trade-off between the values of the output characteristics (fidelity and success probability) can be achieved in the case of extracting smaller photonic state (say, up to 20 photons) in scheme with two BSs and PNR detectors to generate even/odd SCSs of less amplitudes≤ 3. Practical quantum engineering of even/odd large-amplitude≥ 5 high-fidelity≥ 0.99 SCSs by subtraction of large (say, 100 photons) with help of multiple hub currently is a challenge. Progress in the development of more efficient PNR detectors can significantly contribute to the practical quantum engineering of large-amplitude≥ 5 high-fidelity≥ 0.99 even/odd SCSs.

## Methods

**Passing SMSV state through the optical hub.** To trace the influence of the optical hub on the unitary evolution of the initial SMSV state in Fig. 1, it is worth using the system of transformations on the creation operators imposed by each of $BS_i$: $a_0^+ \to t_i a_0^+ - r_i a_i^+$, $a_i^+ \to r_i a_0^+ + t_i a_i^+$, where $a_0^+$ is the creation operator of the light field propagating in 0 mode and $a_i^+$ is the creation operator of $i$ light field generated by $BS_i$. The transformations are the basis to derive output entangled state. Finally, output entangled state produced by series of $k$ BSs over SMSV inputted into mode 0 is given by

$$BS_{0k}BS_{0k-1}\ldots BS_{0i}\ldots BS_{01}(|SMSV\rangle_0|0\rangle_1|0\rangle_2\ldots|0\rangle_i\ldots|0\rangle_k) = \frac{1}{\sqrt{\cosh s}}$$

$$\sum_{n_1=0}^{\infty}\cdots\sum_{n_1=0}^{\infty}\cdots\sum_{n_k=0}^{\infty}(-1)^{n_1+n_2+\cdots+n_k}C_{n_1,n_2,\ldots,n_k}^{(0,0,\ldots,0)}\left|\Psi_{n_1,\ldots,n_i,\ldots,n_k}^{(0,0,\ldots,0)}\right\rangle_0|n_1\rangle_1\ldots|n_i\rangle_i\ldots|n_k\rangle_k, \quad (22)$$

where $BS_{0i}$ means the beam splitter operator mixing modes 0 and $i$. Here, the CV states with even total number $N_k = n_1 + n_2 + \cdots + n_k = 2m_N$ of photons reflected to ancillary measurement modes are presented by

$$\left|\Psi_{n_1,\ldots,n_i,\ldots,n_k}^{(0,0,\ldots,0)}\right\rangle = \left|\Psi_{2m_N}^{(0,0,\ldots,0)}\right\rangle = \frac{1}{\sqrt{Z^{(2m_N)}(y_k)}}\sum_{n=0}^{\infty}\frac{y_k^n}{\sqrt{(2n)!}}\frac{(2(n+m_N))!}{(n+m_N)!}|2n\rangle, \quad (23)$$

while the CV states with odd total number $N_k = n_1 + n_2 + \cdots + n_k = 2m_N + 1$ of photons redirected to ancillary measuring modes are given by

$$\left|\Psi_{n_1,\ldots,n_i,\ldots,n_k}^{(0,0,\ldots,0)}\right\rangle = \left|\Psi_{2m_N+1}^{(0,0,\ldots,0)}\right\rangle = \sqrt{\frac{y_k}{Z^{(2m_N+1)}(y_k)}}\sum_{n=0}^{\infty}\frac{y_k^n}{\sqrt{(2n+1)!}}\frac{(2(n+m_N+1))!}{(n+m_N+1)!}|2n+1\rangle. \quad (24)$$

Amplitudes of the entangled state in Eq. (2) are given by

$$C_{n_1,n_2,\ldots,n_k}^{(0,0,\ldots,0)} = \prod_{l=1}^{k}\left(\frac{1-t_l^2}{t_l^2}\right)^{\frac{n_l}{2}}\frac{y_l^{\frac{n_l}{2}}}{\sqrt{n_l!}}\begin{cases}\sqrt{Z^{(2m_N)}(y_k)}, & \text{if } N_k = 2m_N \\ \sqrt{Z^{(2m_N+1)}(y_k)}, & \text{if } N_k = 2m_N + 1\end{cases}. \quad (25)$$

Here by definition, the following function $Z \equiv Z(y) = Z^{(0)} = 1/\sqrt{1-4y^2}$ and its derivative $Z^{(m)} = d^m Z/dy^m$ with respect to the parameter $y = t^2\tanh s/2$ determined through the experimental parameters $(t,s)$ are introduced. For the formulas (23-25), argument of the function $Z(y)$ and $y$ should be specified. If only one beam splitter is used in Figure 1, then the parameter becomes $y_1 = t_1^2 \tanh s/2$, that is, it differs from the original argument $y_0$ by the value $t_1^2$ i.e. $y_1 = t_1^2 y_0$. By definition, the parameter $y_1$ can also take values in the range $0 < y_1 < 0.5$ in the case of $s > 0$. Note that the case $y_1 = 0$ is realized either in the absence of the SMSV at the input to the BS ($s = 0$) or in the case of reflection of all photons into the second auxiliary mode that is, when $t_1 = 0, r_1 = 1$, while the case of $y_1 = 0.5$ can only appears in the non-physical case of $s \to \infty$ and $t_1 = 1$. Accordingly, the function $Z(y)$ in Eqs. (23-25) is redefined to be $Z(y_1) = 1/\sqrt{1 - 4y_1^2}$ for the hub with one element. The passage of the original SMSV through $i-$BS leads to a change of the input parameter $y_0$ to $y_i$ as $y_0 \to y_i = (t_1^2 t_2^2 \dots t_i^2) \tanh s/2 = t_1^2 t_2^2 \dots t_i^2 y_0 = t_i^2 y_{i-1}$ for the output CV states in Eq. (23,24) in zeroth mode. It is easy to observe that the parameter $y_{i-1}$ acquires an additional reducing factor $t_i^2$ after CV state has passed the next $i$ BS i.e. $y_i = t_i^2 y_{i-1}$. Finally, output state's parameter $y_k$ used in Eqs. (23-25) becomes $y_k = (t_1^2 t_2^2 \dots t_i^2 \dots t_k^2) \tanh s/2 = t_1^2 t_2^2 \dots t_i^2 \dots t_k^2 y_0$ after the initial SMSV goes through all $k$ beam splitters. Thus, the action of the system of $k$ successive BSs causes a decrease in the initial parameter $y_0$ in $t_1^2 t_2^2 \dots t_i^2 \dots t_k^2$ times, i.e. $y_k/y_0 = t_1^2 t_2^2 \dots t_i^2 \dots t_k^2$. Now, the final function $Z$ that determines the normalization factor of the CV states in Eqs. (23,24) also depends on $y_k$, i.e. $Z(y_k) = 1/\sqrt{1 - 4y_k^2}$. The limiting values of the parameter $y_k$ can be taken in the case of either $s = 0$ or $t_i = 0$ leading to $y_k = 0$ or in the case of $s \to \infty$, $t_1 = t_2 = \dots = t_i = \dots t_k = 1$ resulting in $y_i = 0.5$ which are not of interest.

**ACKNOWLEDGEMENT**
MSP, SAP and SPK are supported by the Ministry of Science and Higher Education of the Russian Federation on the basis of the FSAEIHE SUSU (NRU) (Agreement No. 075-15-2022-1116).

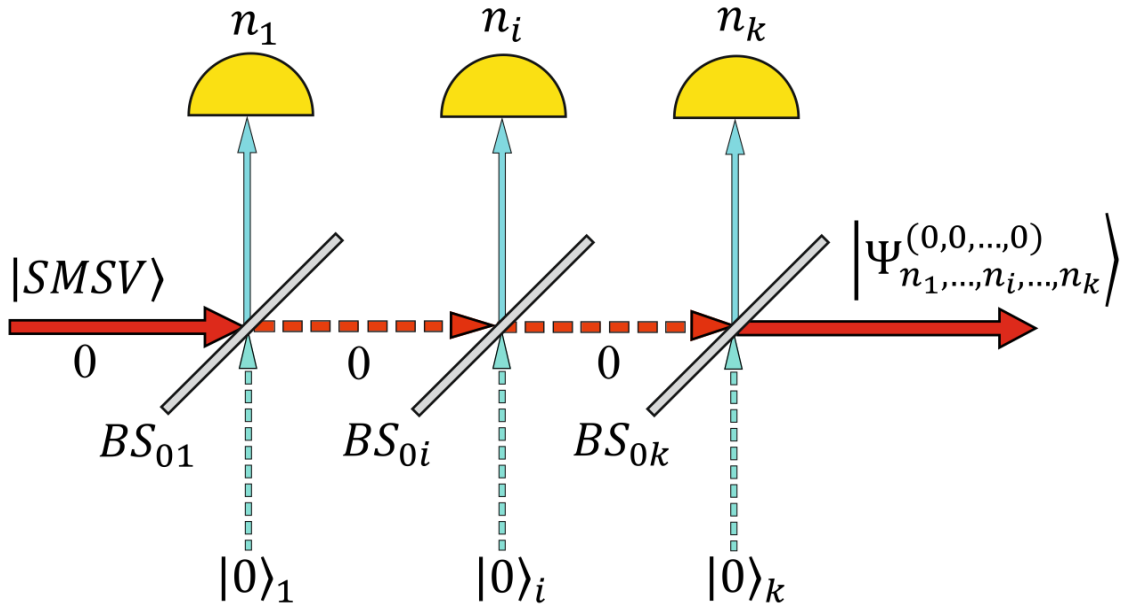

**Figure 1** Optical scheme used to shape even/odd SCS of large amplitude with fidelity over 0.99. It consists of a sequence of $k$ beam splitters with transmittance coefficient $t_i$ ($i = 1,2,\ldots,k$) following one after another through which the original SMSV state with squeezing amplitude $s$ passes forming the entangled state in Eq. (2). Part of the reflected photons ($n_1, n_2, \ldots, n_k$) is simultaneously measured in auxiliary modes resulting in either even CV (Eq. (3)) in the case of $n_1 + n_2 + \cdots + n_k = 2m_N$ or odd CV (Eq. (4)) heralded states provided that $n_1 + n_2 + \cdots + n_k = 2m_N + 1$. A conditional state can approximate either an even or an odd SCS under certain values of one parameter $y_k$ defined by $s$ and $t_i$ ($i = 1,2,\ldots,k$).

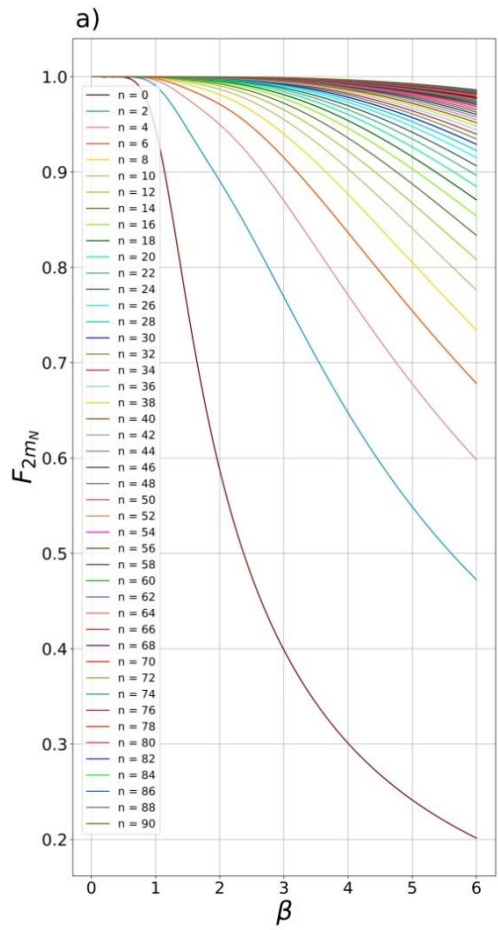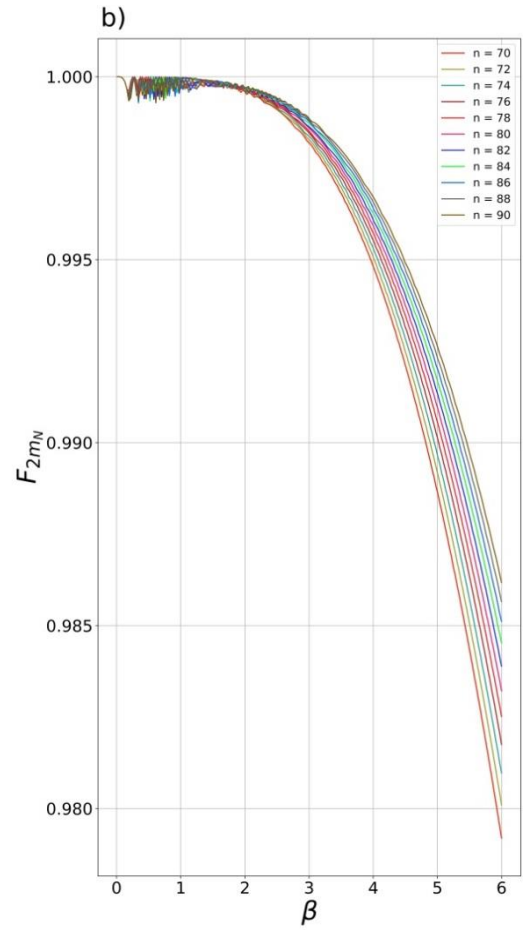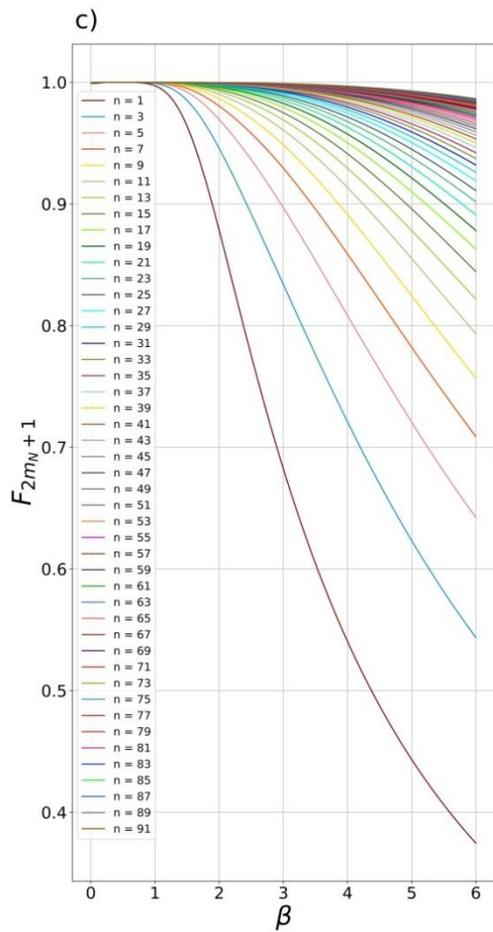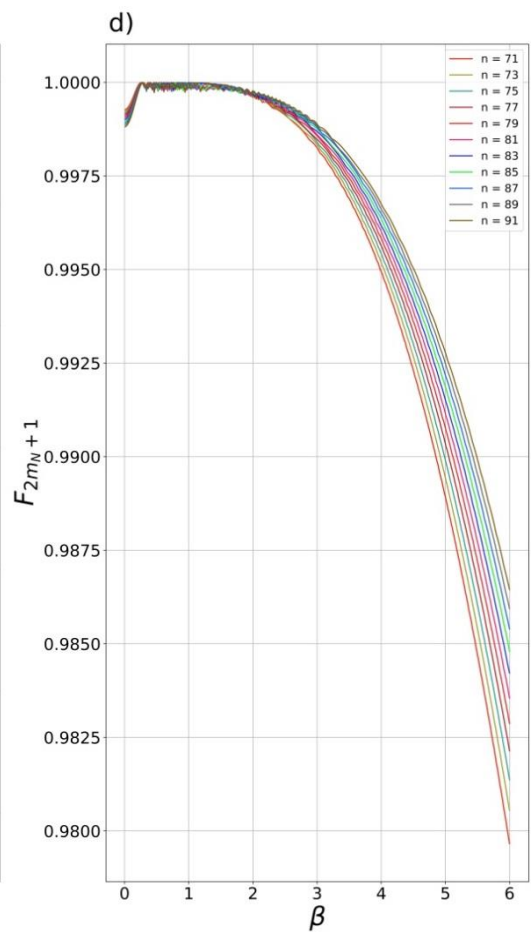
16

**Figure 2(a-d)** Dependence of the even (a,b) $F_{2m}$ and odd (c,d) $F_{2m+1}$ fidelities between $2m, 2m+1$ −heralded CV states of definite parity and even/odd SCSs on the SCS amplitude $\beta$. The more photons $n$ is measured in auxiliary modes, the higher fidelities $F_n$ of the generated states is observed. Dependences of the fidelities of higher-order CV states with $n$ from 70 up to 91 depending on $\beta$ are separately shown in subfigures (b) и (d). The dependences on subfigures (b) и (d) allow for one to observe generation of even/odd SCSs with an amplitude greater than 5 with fidelity exceeding 0.99.

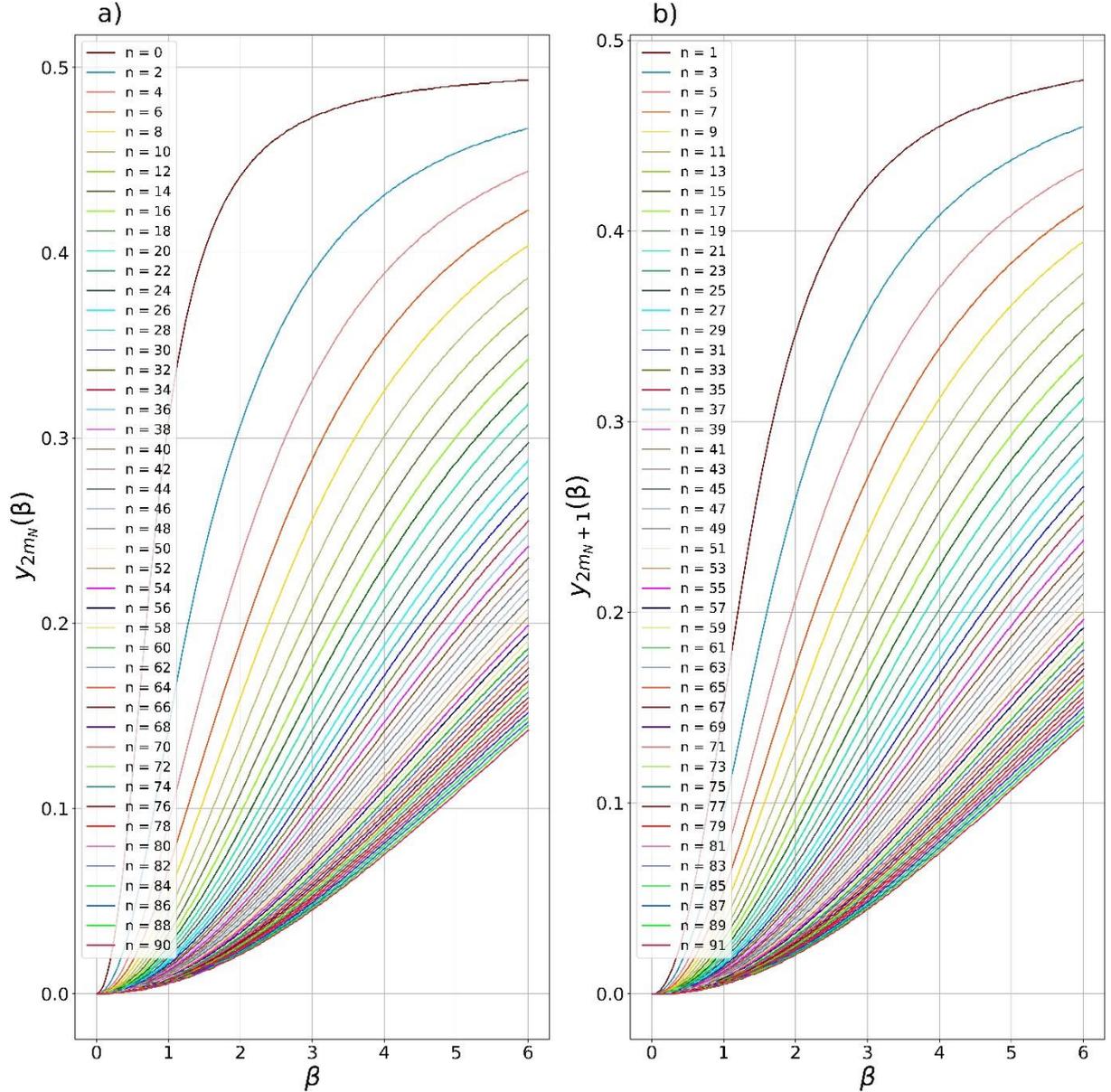

**Figure 3(a,b)** $2m, 2m+1$ −heralded states in Eqs. (3,4) depend solely on one CV parameter $0 \leq y_k \leq 0.5$. The plots demonstrate the dependencies of the parameter (a) $y_{2m_N}(\beta)$ and (b) $y_{2m_N+1}(\beta)$ that provide the fidelities in Fig. 2 if either $y_k = y_{2m_N}(\beta)$ or $y_k = y_{2m_N+1}(\beta)$. The larger the number $n$ of the extracted photons, the smaller the value of the parameters $y_{2m_N}(\beta)$ and $y_{2m_N+1}(\beta)$, moreover accompanied by an increase in the fidelity of the conditional states.

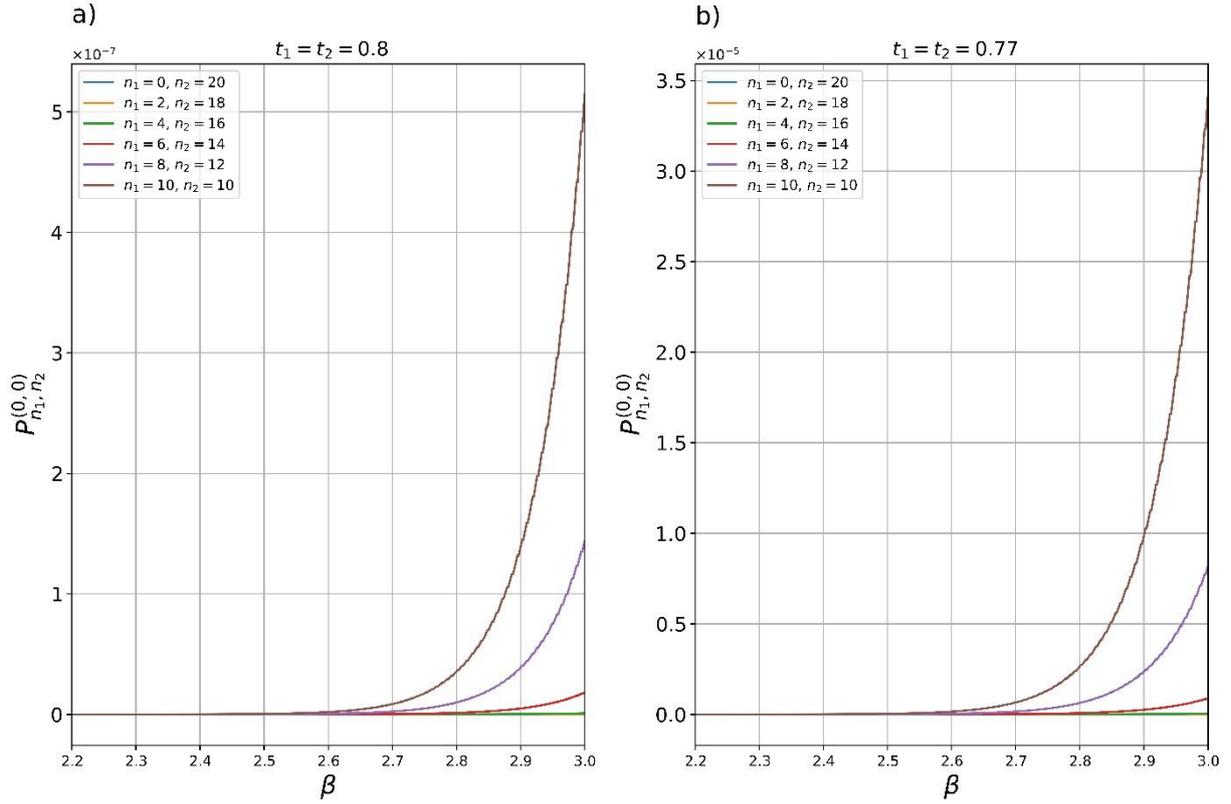

**Figure 4(a,b)** Plots of the success probabilities $P_{n_1,n_2}^{(0,0)}(y_{20}(\beta))$ (Eq. (13)) to generate even SCSs with the fidelity presented in Figure 2(a) on output of the hub with two BSs and PNR detectors in the dependency on the SCS amplitude $\beta$. The plots are constructed for different BSs transmittance coefficients: a) $t_1 = t_2 = 0.8$ and b) $t_1 = t_2 = 0.77$. Detection of only even photon states $n_1 = 2m_1$ and $n_2 = 2m_2$ is considered. The maximum possible probability $P_{10,10}^{(0,0)}(y_{20}(\beta))$ is observed when $n_1 = 10$ and $n_2 = 10$. Moreover, the probability is several orders of magnitude higher than the probability $P_{0,20}^{(0,0)}(y_{20}(\beta))$, which almost coincides with the probability $P_{20}^{(0)}(y_{20}(\beta))$ of generating SCS in a scheme with one BS and PNR detector $\left(P_{0,20}^{(0,0)}(y_{20}(\beta)) \cong P_{20}^{(0)}(y_{20}(\beta))\right)$. Transmission of more photons into the measurement modes (b) makes it possible to increase the success probability by almost two orders of magnitude.

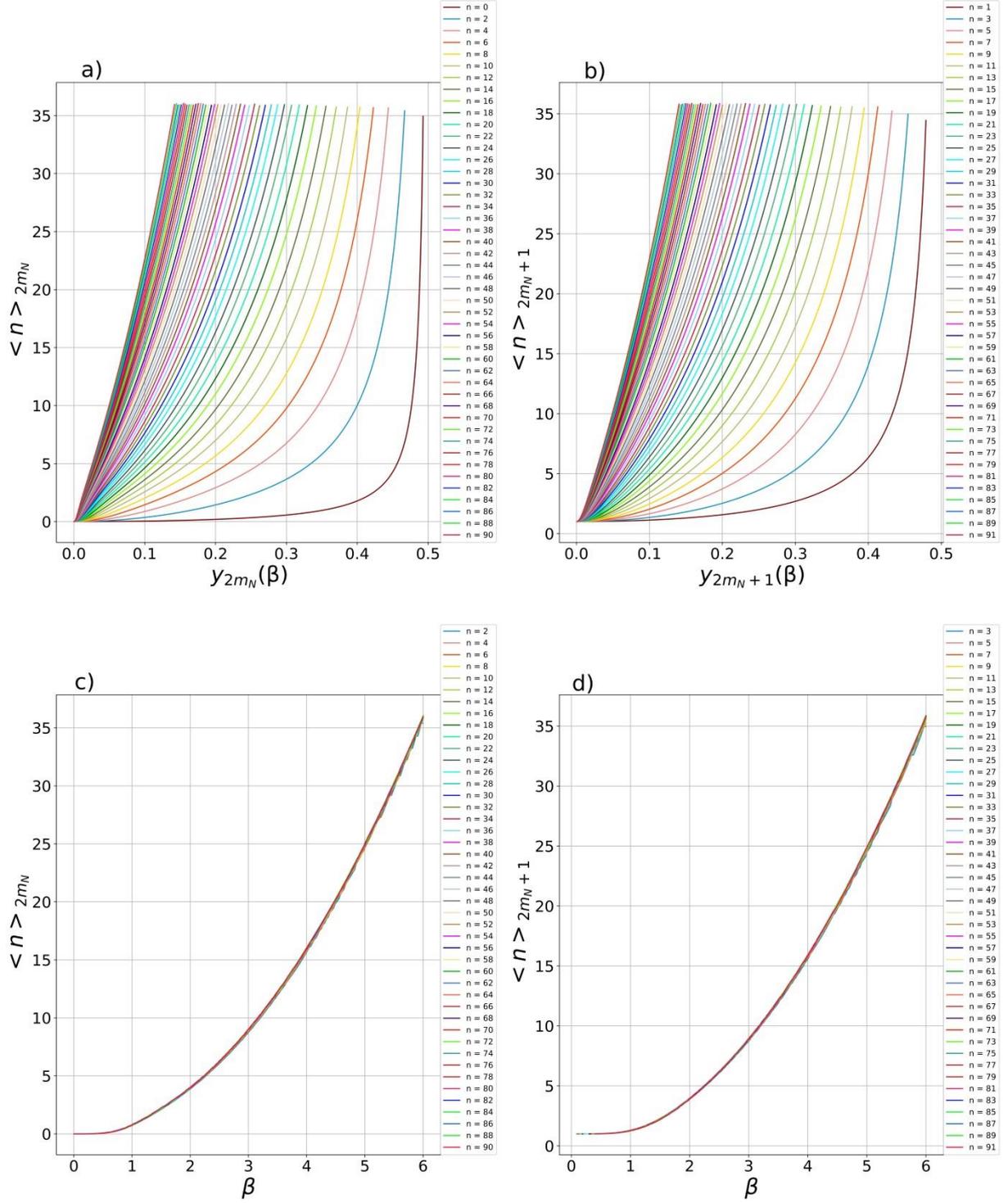

**Figure 5(a-d)** Dependence of average number of photons a) $\langle n \rangle_{2m_N}$ and b) $\langle n \rangle_{2m_N+1}$ in generated CV states on parameters $y_{2m_N}(\beta)$ and $y_{2m_N+1}(\beta)$, respectively, which provide the maximum fidelity of the states with even/odd SCSs. When constructing c) $\langle n \rangle_{2m_N}$ and d) $\langle n \rangle_{2m_N+1}$ depending on the corresponding SCS amplitude $\beta$, the curves coincide with each other which confirms the rule for the SCSs i.e. $\langle n \rangle_{2m_N} \approx \beta^2$ and $\langle n \rangle_{2m_N+1} \approx \beta^2$.